\documentclass[orivec,conference]{llncs}
\usepackage[cmex10]{amsmath}
\usepackage{stmaryrd}
\usepackage{amsfonts}
\usepackage{amssymb}
\usepackage{bbold}
\usepackage{wrapfig}
\usepackage{graphicx}
\usepackage{enumerate}
\usepackage{multirow}
\usepackage{xspace}
\usepackage{cancel}
\usepackage{enumitem}
\usepackage[hidelinks]{hyperref}
\usepackage{esvect}
\usepackage{caption}
\usepackage{xcolor,colortbl}
\newcommand{\Dpqe}{\mbox{$\mi{DS}$-$\mi{PQE}$}\xspace}
\newcommand{\dpqe}{\mbox{$\mi{ds}$-$\mi{pqe}$}\xspace}

\newcommand{\olds}[4]{\mbox{$(\prob{#1}{#2},\pnt{#3}\,)\rightarrow #4$}}

\newcommand{\Oods}[3]{\mbox{$(\pnt{#1},#2)\rightarrow #3$}}
\newcommand{\oDs}[3]{\mbox{$(#1,#2)\rightarrow #3$}}

\newcommand{\oods}[2]{\mbox{\pnt{#1}~$\rightarrow #2$}}
\newcommand{\ods}[2]{\mbox{$#1 \rightarrow #2$}}
\newcommand{\bm}[1]{{\mbox{\boldmath $#1$}}}
\newcommand{\Bm}[1]{{\boldmath $#1$}}

\newcommand{\imp}{\Rightarrow} 
\newcommand{\ie}{i.e.,\xspace}
\newcommand{\eg}{e.g.,\xspace}
\newcommand{\pnt}[1]{\mbox{$\vv{#1}$}\xspace}
\newcommand{\ppnt}[2]{\mbox{$\vv{#1}\!_{#2}$}}

\newcommand{\pent}[2]{\mbox{$\vv{#1}#2$}}

\newcommand{\cof}[2]{\mbox{$#1_{\vec{#2}}$}}

\newcommand{\V}[1]{\mbox{$\mathit{Vars}(#1)$}}
\newcommand{\Va}[1]{\mbox{$\mi{Vars}(\vec{#1})$}}

\newcommand{\s}[1]{\mbox{$\{#1\}$}}

\newcommand{\nGz}[2]{$G_{non-\{z\}}$}

\newcommand{\prr}[1]{\mi{Prev}(\boldsymbol{q})}

\newcommand{\mi}[1]{\mathit{#1}}
\newcommand{\ti}[1]{\textit{#1}}
\newcommand{\tb}[1]{\textbf{#1}}

\newcommand{\Tt}{\>\>}
\newcommand{\Sub}[2]{\mbox{$\mi{#1}\!_{\mi{#2}}$}}

\newcommand{\prob}[2]{\mbox{$\exists{#1} [#2]$}}

\newcommand{\Comment}[1]{}

\newcommand{\aps}[1]{\mbox{$\mathbb{#1}$}}

\begin{document}

\title{On Efficient Algorithms For Partial Quantifier Elimination}


\author{Eugene Goldberg}
\institute{\email{eu.goldberg@gmail.com}}

\maketitle

\begin{abstract}
Earlier, we introduced Partial Quantifier Elimination (PQE). It is a
$\mathit{generalization}$ of regular quantifier elimination where one
can take a $\mathit{part}$ of the formula out of the scope of
quantifiers. We apply PQE to CNF formulas of propositional logic with
existential quantifiers. The appeal of PQE is that many problems like
equivalence checking and model checking can be solved in terms of PQE
and the latter can be very efficient. The main flaw of current PQE
solvers is that they do not $\mathit{reuse}$ learned information. The
problem here is that these PQE solvers are based on the notion of
clause redundancy and the latter is a $\mathit{structural}$ rather
than $\mathit{semantic}$ property. In this paper, we provide two
important theoretical results that enable reusing the information
learned by a PQE solver.  Such reusing can dramatically boost the
efficiency of PQE like conflict clause learning boosts SAT solving.
\end{abstract}


\section{Introduction}
\label{sec:intro}

Partial Quantifier Elimination (PQE) is a generalization of regular
quantifier elimination (QE). In PQE one can take a \ti{part} of the
formula out of the scope of quantifiers hence the name \ti{partial}.
The appeal of PQE is twofold.  First, many old problems like
equivalence checking, model checking and SAT and new problems like
property generation~\cite{cav23} can be solved in terms of PQE.
Second, PQE can be dramatically more efficient than QE.

PQE is defined as follows~\cite{hvc-14}. Let $F(X,Y)$ be a
\ti{propositional} formula in conjunctive normal
form\footnote{Given a CNF formula $F$ represented as the conjunction of clauses
$C_0 \wedge \dots \wedge C_k$, we will also consider $F$ as
the \ti{set} of clauses \s{C_0,\dots,C_k}.
} (CNF) where $X,Y$ are sets of
variables. Let $G$ be a subset of clauses of $F$.  Given a formula
\prob{X}{F}, the PQE problem is to find a quantifier-free formula
$H(Y)$ such that $\prob{X}{F}\equiv H\wedge\prob{X}{F \setminus G}$.
In contrast to \ti{full} QE, only the clauses of $G$ are taken out of
the scope of quantifiers.  In this paper, we consider PQE for formulas
with only \ti{existential} quantifiers. We will refer to $H$ as a
\ti{solution} to PQE.  Note that QE is just a special case of PQE
where $G = F$ and the entire formula is unquantified. A key role in
PQE solving is played by \ti{redundancy based reasoning}: to take a
set of clauses $G$ out of \prob{X}{F(X,Y)}, one essentially needs to
find a formula $H(Y)$ that makes $G$ \ti{redundant} in $H \wedge
\prob{X}{F}$.  Redundancy based reasoning helps to explain the
efficiency of PQE over QE. If $G$ is small, finding $H$ that makes $G$
redundant in $H \wedge \prob{X}{F}$ is much easier than making
redundant the \ti{entire} formula $F$.

The current PQE algorithms implement redundancy based reasoning via
the machinery of D-sequents~\cite{fmcad12,fmcad13}. A D-sequent is a
record stating the redundancy of a clause $C$ in the current formula
\prob{X}{F} in subspace \pnt{q}. Here \pnt{q} is an assignment to
variables of $F$ and ``D'' stands for ``Dependency''.  The redundancy
of $C$ above also holds in any formula \prob{X}{F^*} where $F^*$ is
obtained from $F$ by adding clauses implied by $F$.

A PQE solver branches on variables of $F$ until it reaches the
subspace where every clause of $G$ can be trivially proved or made
redundant.  At this point, so called atomic D-sequents are
generated. D-sequents derived in different branches can be resolved
similarly to clauses. The objective of a PQE solver is to derive, for
each clause $C \in G$, the ``global'' D-sequent stating the redundancy
of $C$ in the final formula \prob{X}{F} in the \ti{entire} space. This
D-sequent is obtained by resolving D-sequents generated in
subspaces. The derivation of global D-sequents, in general, requires
adding to \prob{X}{F} new clauses. The unquantified clauses (\ie those
depending only $Y$) that have been added to the initial formula
\prob{X}{F} form a solution $H(Y)$ to the PQE problem above.

A major flaw of current PQE solvers is that they \ti{do not reuse}
D-sequents (in contrast to SAT-solvers that derive their power from
reusing conflict clauses). The reason for that is as follows.  A
conflict clause expresses a local \ti{inconsistency}, the latter being
a \ti{semantic} property. This simplifies reusing conflict clauses in
different contexts. On the other hand, as we argue in
Section~\ref{sec:reuse_pqe}, a D-sequent expresses a local
\ti{unobservability} that is a \ti{structural} rather than semantic
property. This means that reusing D-sequents depends on the context
and so is not trivial.  Importantly, current PQE algorithms repeatedly
derive the same D-sequents (see Appendix~\ref{app:reuse_dseqs}). So,
the lack of reusing D-sequents cripples the performance of PQE
solving. The main contribution of this paper is addressing the issue
of D-sequent reusing. Namely, we present important theoretical results
enabling safe reusing of D-sequents. These results apply to so-called
\ti{extended} D-sequents introduced in~\cite{qe_learn}.

This paper is structured as follows. Section~\ref{sec:basic} gives
some basic definitions. In Section~\ref{sec:pqe_alg} we explain PQE
solving by the example of the PQE algorithm called \Dpqe.  The notion
of extended D-sequents is introduced in Section~\ref{sec:ext_dseqs}.
Section~\ref{sec:reuse_pqe} gives a high-level view of D-sequent
reusing.  In Sections~\ref{sec:sngl_dseq}-\ref{sec:mltpl_dseqs} we lay
a theoretical foundation for D-sequent reusing.  Finally, in
Sections~\ref{sec:bckgr} and~\ref{sec:concl}, we present some
background and make conclusions.

\section{Basic Definitions}
\label{sec:basic}

In this section, when we say ``formula'' without mentioning
quantifiers, we mean ``a quantifier-free formula''.

\begin{definition}
\label{def:cnf}
We assume that formulas have only Boolean variables.  A \tb{literal}
of a variable $v$ is either $v$ or its negation.  A \tb{clause} is a
disjunction of literals. A formula $F$ is in conjunctive normal form
(\tb{CNF}) if $F = C_0 \wedge \dots \wedge C_k$ where $C_0,\dots,C_k$
are clauses. We will also view $F$ as the \tb{set of
clauses} \s{C_0,\dots,C_k}. We assume that \tb{every formula is in
CNF} unless otherwise stated.
\end{definition}
%
%
\begin{definition}
  \label{def:vars} Let $F$ be a formula. Then \bm{\V{F}} denotes the
set of variables of $F$ and \bm{\V{\prob{X}{F}}} denotes
$\V{F}\!\setminus\!X$.
\end{definition}
%
%
\begin{definition}
Let $V$ be a set of variables. An \tb{assignment} \pnt{q} to $V$ is a
mapping $V'\!\rightarrow\!\s{0,1}$ where $V'\!\subseteq V$.  We will
denote the set of variables assigned in \pnt{q} as \bm{\Va{q}}. We
will refer to \pnt{q} as a \tb{full assignment} to $V$ if
$\Va{q}=V$. We will denote as \bm{\pnt{q} \subseteq \pnt{r}} the fact
that a) $\Va{q} \subseteq \Va{r}$ and b) every variable of \Va{q} has
the same value in \pnt{q} and \pnt{r}.
\end{definition}

%
%
\begin{definition}
A literal and a clause are said to be \tb{satisfied}
(respectively \tb{falsified}) by an assignment \pnt{q} if they
evaluate to 1 (respectively 0) under \pnt{q}.
\end{definition}

%
%
\begin{definition}
\label{def:cofactor}
Let $C$ be a clause. Let $H$ be a formula that may have quantifiers,
and \pnt{q} be an assignment to
\V{H}.  If $C$ is satisfied by \pnt{q}, then \bm{\cof{C}{q} \equiv
  1}. Otherwise, \bm{\cof{C}{q}} is the clause obtained from $C$ by
removing all literals falsified by \pnt{q}. Denote by \bm{\cof{H}{q}}
the formula obtained from $H$ by removing the clauses satisfied by
\pnt{q} and replacing every clause $C$ unsatisfied by \pnt{q} with
\cof{C}{q}.
\end{definition}

%
%
\begin{definition}
\label{def:formula-equiv}
Let $G, H$ be formulas that may have existential quantifiers. We say
that $G, H$ are \tb{equivalent}, written \bm{G \equiv H}, if $\cof{G}{q} =
\cof{H}{q}$ for all full assignments \pnt{q} to $\V{G} \cup \V{H}$.
\end{definition}

%
%
\begin{definition}
\label{def:red_cls}
Let $F(X,Y)$ be a formula and $G \subseteq F$ and $G \neq
\emptyset$. The clauses of $G$ are said to be \tb{redundant in}
\bm{\prob{X}{F}} if $\prob{X}{F} \equiv \prob{X}{F \setminus G}$. If
$F \setminus G$ implies $G$, the clauses of $G$ are redundant in
\prob{X}{F} but the opposite is not true.
\end{definition}

%
%
\begin{definition}
  \label{def:Xcls}
Given a formula \prob{X}{F(X,Y)}, a clause $C$ of $F$ is called a
\tb{quantified clause} if \V{C} $\cap~X~\neq~\emptyset$. Otherwise,
$C$ is called \ti{unquantified}.
\end{definition}
%
%
\begin{definition}
 \label{def:pqe_prob} Given a formula \prob{X}{F(X,Y))} and $G$ where
 $G \subseteq F$, the \tb{Partial Quantifier Elimination} (\tb{PQE})
 problem is to find $H(Y)$ such that \Bm{\prob{X}{F}\equiv
   H\wedge\prob{X}{F \setminus G}}.  (So, PQE takes $G$ out of the
 scope of quantifiers.)  The formula $H$ is called a \tb{solution} to
 PQE. The case of PQE where $G = F$ is called \ti{Quantifier
   Elimination} (\tb{QE}).
\end{definition}

%
%
\begin{example}
\label{exmp:pqe_exmp}
Consider formula $F = C_0 \wedge \dots \wedge C_4$ where
$C_0=\overline{x}_2 \vee x_3$, $C_1\!=\!y_0\vee x_2$, $C_2=y_0 \vee
\overline{x}_3$, $C_3\!=\!y_1\!\vee\!x_3$,
$C_4\!=\!y_1\!\vee\!\overline{x}_3$. Let $Y =\s{y_0,y_1}$ and $X =
\s{x_2,x_3}$. Consider the PQE problem of taking out $G$ consisting of
a single clause $C_0$. That is one needs to find $H(Y)$ such that
$\prob{X}{F} \equiv H \wedge \prob{X}{F \setminus \s{C_0}}$. One can
show that\linebreak $\prob{X}{F} \equiv y_0 \wedge \prob{X}{F
  \setminus \s{C_0}}$ (see Subsection~\ref{ssec:pqe_exmp}) \ie $H\!
=\!y_0$ is a solution to this PQE problem.
\end{example}

%
%
\begin{definition}
\label{def:res_of_cls}
Let clauses $C'$,$C''$ have opposite literals of exactly one variable
$w\!\in\!\V{C'}\!\cap\!\V{C''}$.  Then $C'$,$C''$ are called
\tb{resolvable} on~$w$.  Let $C$ be the clause consisting of the
literals of $C'$ and $C''$ minus those of $w$. Then $C$ is said to be
obtained by \tb{resolution} of $C'$ and $C''$ on $w$.
\end{definition}
%
%
\begin{definition}
\label{def:blk_cls}
Let $C$ be a clause of a formula $F$ and $w \in \V{C}$. The clause $C$
is said to be \tb{blocked}~\cite{blocked_clause} in $F$ at the
variable $w$ if no clause of $F$ is resolvable with $C$ on $w$.
\end{definition}

%
%
\begin{proposition}
\label{prop:blk_cls}
Let a clause $C$ be blocked in a formula $F(X,Y)$ with respect to a
variable $x \in X$.  Then $C$ is redundant in \prob{X}{F},
\ie \prob{X}{F \setminus \s{C}} $\equiv$ \prob{X}{F}.
\end{proposition}

The proofs of propositions are given in Appendix~\ref{app:proofs}.

\section{PQE solving}
\label{sec:pqe_alg}

In this section, we briefly describe the PQE algorithm called
\Dpqe~\cite{hvc-14}. Our objective here is just to give an idea of how
the PQE problem can be solved. So, in
Subsection~\ref{ssec:high_level}, we present a high-level description
of this algorithm. (The pseudo-code of \Dpqe is given in
Appendix~\ref{app:pseudo_code}.) Subsections~\ref{ssec:pqe_exmp}
and~\ref{ssec:trg_ch} provide examples of PQE solving.

%
%
\subsection{High-level view}
\label{ssec:high_level}
Like all existing PQE algorithms, \Dpqe uses \ti{redundancy based
  reasoning} justified by the proposition below.
%
%
\begin{proposition}
\label{prop:sol_for_pqe}
  Formula $H(Y)$ is a solution to the PQE problem of taking 
  $G$ out of \prob{X}{F(X,Y)} (\ie $ \prob{X}{F} \equiv H \wedge
  \prob{X}{F \setminus G}$) iff
  \begin{enumerate}
  \item $F \imp H$ and
  \item $H \wedge
    \prob{X}{F} \equiv H \wedge \prob{X}{F \setminus G}$
  \end{enumerate}
\end{proposition}

Thus, to take $G$ out of \prob{X}{F(X,Y)}, it suffices to find a
formula $H(Y)$ implied by $F$ that makes $G$ \ti{redundant} in $H
\wedge \prob{X}{F}$. We will refer to the clauses of $G$ as
\tb{target} ones.
Below, we provide some basic facts about \Dpqe. Since taking out an
unquantified clause is trivial, we assume that $G$ contains only
\ti{quantified} clauses.  \Dpqe finds a solution to the PQE problem
above by branching on variables of $F$.  The idea here is to reach a
subspace \pnt{q} where every clause of $G$ can be easily proved or
made redundant in \prob{X}{F}.  Like a SAT-solver, \Dpqe runs Boolean
Constraint Propagation (BCP). If a conflict occurs in subspace
\pnt{q}, \Dpqe generates a conflict clause $K$ and adds it to $F$ to
\ti{make} clauses of $G$ redundant in subspace \pnt{q}. However, most
frequently, proving redundancy of $G$ in a subspace does not require a
conflict or adding a new clause. Importantly, \Dpqe branches on
unquantified variables, \ie those of $Y$, \ti{before} quantified ones.
So, when \Dpqe produces a conflict clause $K$ from clauses of $F$, the
quantified variables are resolved out \ti{before} unquantified.

If a target clause $C$ becomes unit\footnote{An unsatisfied clause is called \ti{unit} if it has only one
unassigned literal. Due to special decision making of \Dpqe (variables
of $Y$ are assigned before those of $X$), if the target clause $C$
becomes unit, its unassigned variable is always in $X$. 
} in
subspace \pnt{q}, \Dpqe \ti{temporarily} extends the set of target
clauses $G$. Namely, \Dpqe adds to $G$ every clause that is resolvable
with $C$ on its only unassigned variable (denote this variable as
$x$). This is done to facilitate proving redundancy of $C$. If the
added clauses are proved redundant in subspace \pnt{q}, the clause $C$
is blocked at $x$ and so is redundant in subspace \pnt{q}.  The fact
that \Dpqe extends $G$ means that it may need to prove redundancy of
clauses other than those of $G$. The difference is that every clause
of the original set $G$ must be proved redundant \ti{globally} whereas
clauses added to $G$ need to be proved redundant only \ti{locally} (in
some subspaces).

To express the redundancy of a clause $C$ in a subspace \pnt{q}, \Dpqe
uses a record \olds{X}{F}{q}{C} called a \tb{D-sequent}. It states the
redundancy of $C$ in the current formula \prob{X}{F} in subspace
\pnt{q}. This D-sequent also holds in any formula \prob{X}{F^*} where
$F^*$ is obtained from $F$ by adding clauses implied by $F$.  So, we
will \tb{omit} the first parameter and will write the D-sequent above
as \oods{q}{C}. We will assume that it represents the redundancy of
$C$ in subspace \pnt{q} for the \ti{current formula} whatever it
is. For the sake of simplicity, in this section we use ``regular''
D-sequents presented in~\cite{fmcad13}.  In the next section, we will
recall extended D-sequents introduced in~\cite{qe_learn} that will be
used for the rest of the paper.

A D-sequent derived for a target clause when its redundancy is easy to
prove is called \tb{atomic}. D-sequents derived in different branches
can be resolved similarly to clauses\footnote{In the previous papers
  (\eg \cite{fmcad13}) we called the operation of resolving D-sequents
  \ti{join}.}. For every target clause $C$ of the original set $G$,
\Dpqe uses such resolution to eventually derive the D-sequent
\ods{\emptyset}{C}. The latter states that $C$ is globally redundant
in the final formula \prob{X}{F}. At this point \Dpqe terminates.  The
solution $H(Y)$ to the PQE problem found by \Dpqe consists of the
unquantified clauses added to the initial formula \prob{X}{F} to make
$G$ redundant.

%
%
\subsection{An example of PQE solving}
\label{ssec:pqe_exmp}

Here we show how \Dpqe solves
Example~\ref{exmp:pqe_exmp} introduced in Section~\ref{sec:basic}.
Recall that one takes $G = \s{C_0}$ out of \prob{X}{F(X,Y)} where $F =
C_0 \wedge \dots \wedge C_4$ and $C_0=\overline{x}_2 \vee x_3$,
$C_1\!=\!y_0\vee x_2$, $C_2=y_0 \vee \overline{x}_3$,
$C_3\!=\!y_1\!\vee\!x_3$, $C_4\!=\!y_1\!\vee\!\overline{x}_3$ and
$Y=\s{y_0,y_1}$ and $X = \s{x_2,x_3}$.  That is, one needs to find
$H(Y)$ such that $\prob{X}{F} \equiv H \wedge \prob{X}{F \setminus
  \s{C_0}}$.

Assume that \Dpqe picks the variable $y_0$ for branching and first
explores the branch $\pent{q}{'}=(y_0\!=\!0)$. In subspace
\pent{q}{'}, clauses $C_1,C_2$ become unit.  After assigning
$x_2\!=\!1$ to satisfy $C_1$, the clause $C_0$ turns into unit too and
a conflict occurs (to satisfy $C_0$ and $C_2$, one has to assign the
opposite values to $x_3$). After a standard conflict
analysis~\cite{grasp}, the conflict clause $K=y_0$ is obtained by
resolving $C_1$ and $C_2$ with $C_0$. To \ti{make} $C_0$ redundant in
subspace \pent{q}{'}, \Dpqe adds $K$ to $F$.  The redundancy of $C_0$
is expressed by the D-sequent \ods{\pent{q}{'}}{C_0}.  This D-sequent
is an example of an \ti{atomic} D-sequent. It asserts that $C_0$ is
redundant in the current formula \prob{X}{F} in subspace
\pent{q}{'}. More information about D-sequents is given in
Sections~\ref{sec:reuse_pqe}-\ref{sec:mltpl_dseqs}.

Having finished the first branch, \Dpqe considers the second branch:
$\pent{q}{''}\!=\!(y_0\!=\!1)$. Since $C_1$ is satisfied by
\pent{q}{''}, no clause of $F$ is resolvable with $C_0$ on variable
$x_2$ in subspace \pent{q}{''}. Hence, $C_0$ is blocked at variable
$x_2$ and thus redundant in \prob{X}{F} in subspace \pent{q}{''}.  So,
\Dpqe generates the D-sequent \ods{\pent{q}{''}}{C_0}. This D-sequent
is another example of an \ti{atomic} D-sequent. It states that $C_0$
is \ti{already} redundant in \prob{X}{F} in subspace \pent{q}{''}
(without adding a new clause).
Then \Dpqe resolves the D-sequents \ods{(y_0=0)}{C_0} and
\ods{(y_0=1)}{C_0} above on $y_0$. This resolution produces the
D-sequent \ods{\emptyset}{C_0} stating the redundancy of $C_0$ in
\prob{X}{F} in the \ti{entire} space (\ie globally).  Recall that
$\Sub{F}{fin}=K \wedge \Sub{F}{init}$ where \Sub{F}{fin} and
\Sub{F}{init} denote the final and initial formula $F$
respectively. That is $K$ is the only unquantified clause added to
\Sub{F}{init}. So, \Dpqe returns $K=y_0$ as a solution $H(Y)$. The
clause $K$ is indeed a solution since it is implied by \Sub{F}{init}
and $C_0$ is redundant in $K \wedge \prob{X}{\Sub{F}{init}}$. So both
conditions of Proposition~\ref{prop:sol_for_pqe} are met and thus
$\prob{X}{\Sub{F}{init}} \equiv y_0 \wedge \prob{X}{\Sub{F}{init}
  \setminus \s{C_0}}$.

%
%
\subsection{An example of adding temporary targets}
\label{ssec:trg_ch}

Let $F=C_0 \wedge C_1 \wedge C_2 \wedge \dots$ where $C_0=y_0 \vee
x_1$, $C_1 = \overline{x}_1 \vee x_2 \vee x_3$, $C_2 = \overline{x}_1
\vee \overline{x}_2 \vee \overline{x}_3$. Let $C_1$ and $C_2$ be the
only clauses of $F$ with the literal $\overline{x}_1$. Consider the
problem of taking formula $G$ out of \prob{X}{F(X,Y)} where $G =
\s{C_0}$ (we assume that $y_0 \in Y$ and $x_1,x_2,x_3 \in X$). Suppose
\Dpqe explores the branch $\pnt{q} = (y_0=0)$. In the subspace
\pnt{q}, the target clause $C_0$ turns into the unit clause $x_1$. In
this case, \Dpqe adds to the set of targets $G$ the clauses $C_1$ and
$C_2$ \ie the clauses that are resolvable with $C_0$ on $x_1$.

The idea here is to facilitate proving redundancy of $C_0$ in subspace
\pnt{q}. Namely, if $C_1$ and $C_2$ are proved redundant in subspace
\pnt{q}, the target $C_0$ becomes \ti{blocked} and hence redundant in
subspace \pnt{q}. Clauses $C_1,C_2$ are added to $G$ only in subspace
\pnt{q} \ie \ti{temporarily}. As soon as \Dpqe backtracks from this
subspace, $C_1,C_2$ are removed from $G$.

\section{Extended D-sequents}
\label{sec:ext_dseqs}
Our results on D-sequent reusing presented in the next three sections
are formulated in terms of extended D-sequents. The latter were
introduced in~\cite{qe_learn}. In Subsection~\ref{ssec:motiv}, we
explain the motivation for extending the notion of a D-sequent.
Subsection~\ref{ssec:ext_atom} presents extended \ti{atomic}
D-sequents.  In Subsection~\ref{ssec:ext_res} we introduce the
\ti{resolution} of extended D-sequents. (In
Subsection~\ref{ssec:pqe_exmp} we already gave examples of atomic
D-sequents and resolution of in terms of regular D-sequents.)  For the
sake of clarity, we will describe the machinery of extended D-sequents
by the example of \Dpqe.  However, the contents of this and the next
three sections applies to any branching algorithm based on extended
D-sequents.
%
%
\subsection{Motivation for extending the notion of a D-sequent}
\label{ssec:motiv}
Example~\ref{exmp:circ_reason} below shows that an indiscriminate
reusing of ``regular'' D-sequents introduced in~\cite{fmcad13} can
lead to \ti{circular reasoning}. Definition~\ref{def:extension}
extends the notion of a D-sequent~\cite{qe_learn} to avoid circular
reasoning.
%
%
\begin{example}
\label{exmp:circ_reason}
Let formula \prob{X}{F(X,Y)} have clauses
$C_0\!=\!x_0\!\vee\!x_2\!\vee x_3$ and \linebreak $C_1\!=\!x_1\!\vee
x_2\!\vee\!x_3$. Suppose in some branch $(\dots,x_0\!=\!0,\dots)$
\Dpqe derived the D-sequent \ods{(x_0=0)}{C_1} due to the fact that
$C_0$ implies $C_1$ when $x_0=0$. (So $C_1$ is redundant in subspace
$x_0\!=\!0$.) Suppose in another branch $(\dots,x_1\!=\!0,\dots)$
\Dpqe derived the D-sequent \ods{(x_1=0)}{C_0} due to the fact that
$C_1$ implies $C_0$ when $x_1\!=\!0$. Suppose \Dpqe explores the
branch $\pnt{q}=(x_0\!=\!0,x_1\!=\!0)$. Although these D-sequents hold
in subspace \pnt{q} \ti{individually}, applying them \ti{together}
means claiming that \ti{both} $C_0$ and $C_1$ are redundant in
subspace \pnt{q}. This is the wrong conclusion derived by circular
reasoning $\blacksquare$
\end{example}
%
%
\begin{definition}
\label{def:extension}
Given a formula \prob{X}{F}, an \tb{extended} D-sequent is a record
\Oods{q}{U}{C} stating that a clause $C$ is redundant in formula
\prob{X}{F} in subspace \pnt{q}.  The set $U$ specifies the set of
quantified clauses \ti{used} to prove redundancy of $C$ in subspace
\pnt{q}.  The assignment
\pnt{q} is called the \tb{conditional} of this D-sequent. The
set $U$ is called its \tb{construction set}.
\end{definition}

The idea of~\cite{qe_learn} is to reuse an extended D-sequent
\Oods{q}{U}{C} \ti{only} if all the clauses of $U$ are still present
in \prob{X}{F} in subspace \pnt{q}. Consider
Example~\ref{exmp:circ_reason} above. Note that $C_0$ was used to
prove $C_1$ redundant and vice versa. So, the extension of the
D-sequents mentioned there looks like \Oods{q}{U'}{C_1} where
$U'=\s{C_0}$ and \Oods{q}{U''}{C_0} where $U'' = \s{C_1}$.  (More
details about atomic extended D-sequents are given in the next
subsection.) If, for instance, one applies \Oods{q}{U'}{C_1} to remove
$C_1$ from \prob{X}{F} in subspace \pnt{q}, the D-sequent
\Oods{q}{U''}{C_0} \ti{cannot} be reused because $C_1\!\in\!U''$ has
been removed. Thus, one avoids circular reasoning.

\subsection{Extended atomic D-sequents}
\label{ssec:ext_atom}

%
%
\begin{definition}
\label{def:atomic}
Suppose \Dpqe takes a set of clauses $G$ out of \prob{X}{F}. Suppose
\Dpqe entered a subspace \pnt{q} and $C$ is a clause of $G$.  \Dpqe
derives a D-sequent called \tb{atomic} if one of the conditions below
is met
 \begin{itemize}
 \item \pnt{q} \ti{satisfies} $C$
 \item \cof{C}{q} is \ti{implied} by \cof{C'}{q} where $C'$ is another
  clause of $F$
 \item \cof{C}{q} is \ti{blocked} in \cof{F}{q} at an unassigned
   variable $x \in X$
 \end{itemize}
\end{definition}

Now we describe how extended atomic D-sequents are built. Assume \Dpqe
enters a subspace \pnt{q} where $C$ is \tb{satisfied}.  Then \Dpqe
derives a D-sequent \Oods{r}{U}{C} where $U = \emptyset$ and \pnt{r}
is a smallest subset of \pnt{q} still satisfying $C$.
Now, assume that $C$ is \tb{implied} by another clause $C'$ of $F$ in
subspace \pnt{q}. Then \Dpqe derives a D-sequent \Oods{r}{U}{C} where
\pnt{r} is the smallest subset of \pnt{q} such that $C$ is still
implied by $C'$ in subspace \pnt{r}. Here $U = \s{C'}$ because $C$ is
redundant due to the presence of $C'$.

Finally, assume that $C$ is \tb{blocked} in subspace \pnt{q} at a
variable $x \in X$. Then every clause of $F$ resolvable with $C$ on
$x$ is either satisfied by \pnt{q} or proved redundant in subspace
\pnt{q}.  Let $C_0,\dots,C_k$ be the clauses of $F$ resolvable with
$C$ on $x$ that are proved redundant in subspace \pnt{q}.  Let
\oDs{\ppnt{q}{0}}{U_0}{C_0},$\dots$,\oDs{\ppnt{q}{k}}{U_k}{C_k}, be
the D-sequents stating the redundancy of those clauses where
$\ppnt{q}{i} \subseteq \pnt{q}$, $i=0,\dots,k$. Then \Dpqe derives a
D-sequent \Oods{r}{U}{C} where \pnt{r} is a smallest subset of \pnt{q}
such that $C$ is still blocked at $x$ in subspace \pnt{r}.  The
construction set $U$ equals $U_0 \cup \dots \cup U_k$.

\begin{example}
\label{exmp:shorten}
Let $C=x_0 \vee x_1$ be a target clause and \Dpqe enters subspace
$\pnt{q}\!=\!(x_0\!=\!1,x_1\!=\!0,\dots)$ where $C$ is satisfied.
Then \Dpqe derives the D-sequent \Oods{r}{U}{C} where
$\pnt{r}\!=\!(x_0\!=\!1)$ is the smallest subset of \pnt{q} that still
satisfies $C$. The construction set $U$ is empty.
\end{example}

\subsection{Resolution of extended D-sequents}
\label{ssec:ext_res}
%
%
\begin{definition}
\label{def:res_of_dseqs}
 Let \oDs{\pent{q}{'}}{U'}{C} and
\oDs{\pent{q}{''}}{U''}{C} be extended D-sequents.
Let the conditionals \pent{q}{'} and \pent{q}{''} have the following
property: precisely one variable $v$ of $\V{\pent{q}{'}}
\cap \V{\pent{q}{''}}$ has different values in \pent{q}{'} and
\pent{q}{''}. Let \pnt{q} be equal to $\pent{q}{'} \cup \pent{q}{'}$
minus the assignments to $v$ and $U\!=\!U' \cup U''$.  The
D-sequent \Oods{q}{U}{C} is said to be obtained by the \tb{resolution}
of the D-sequents above on $v$.
\end{definition}
%
%
\begin{example}
\label{exmp:resol}
Consider the D-sequents \oDs{\pent{q}{'}}{U'}{C} and
\oDs{\pent{q}{''}}{U''}{C} where
$\pent{q}{'}\!=\!(x_0\!=\!0,x_1\!=\!1,x_2\!=\!1)$ and
$\pent{q}{''}\!=\!(x_0\!=\!0,x_1\!=\!0,x_3\!=\!0)$.  The conditionals
\pent{q}{'} and \pent{q}{''} have exactly one variable that is
assigned differently in them (namely $x_1$). So, these D-sequents can
be resolved on $x_1$. The result of resolution is the D-sequent
\Oods{q}{U}{C} where $\pnt{q}\!=\!(x_0\!=\!0,x_2\!=\!1,x_3\!=\!0)$ and
$U = U' \cup U''$.
\end{example}

\section{A High-Level View Of D-sequent Reusing}
\label{sec:reuse_pqe}
The main flaw of current PQE solvers is that they do not reuse learned
information.  In Appendix~\ref{app:reuse_dseqs} we show experimentally
that due to the lack of D-sequent reusing \Dpqe generates the same
D-sequents over and over again. This suggests that the performance of
PQE solvers like \Dpqe can be dramatically improved by reusing learned
D-sequents.  The same applies to any algorithms based on the machinery
of D-sequents. In this section, we give a high-level view of D-sequent
reusing and in the following two sections we describe two important
theoretical results facilitating such reusing.  As before, for the
sake of clarity, we describe D-sequent reusing by the example of
\Dpqe.

\subsection{Difference between reusing conflict clauses and  D-sequents}
Consider the SAT problem specified by \prob{X}{F(X)}. A typical SAT
solver enumerates only subspaces where the formula is unsatisfiable
(\ie locally inconsistent).  As soon as a satisfying assignment is
found, this SAT solver terminates. Inconsistency is a \ti{semantic}
property. That is, if a formula $R$ is unsatisfiable and $R' \equiv
R$, then formula $R'$ is unsatisfiable too. For this reason, one can
easily easily reuse a conflict clause (specifying local inconsistency)
in different contexts.

Now consider the PQE problem of taking a set of clauses $G$ out of
\prob{X}{F(X,Y)}. A PQE algorithm solves this problem by deriving
D-sequents stating redundancy of clauses of $G$ in subspaces. One can
relate such redundancy with local \ti{unobservability}\footnote{In
  general, when solving a problem involving a quantified formula one
  has to enumerate subspaces where this formula is unsatisfiable
  and those where it is \ti{satisfiable}. So, one cannot get away with
  just recording local inconsistencies as it is done in SAT solving.}.
We mean that if a clause $C$ is redundant in \prob{X}{F} in subspace
\pnt{q}, the presence of $C$ does not affect the result of solving the
PQE problem at hand in this subspace. So, $C$ is ``unobservable''
there. (Appendix~\ref{app:unobs} gives an example of employing
D-sequents to express the unobservability of a subcircuit in a
subspace.)  Note that redundancy and hence unobservability is a
\ti{structural} rather than semantic property. Namely, if a clause $B$
is redundant in formula $R$ and $R' \equiv R$, this \ti{does not}
necessarily mean that $B$ is redundant in $R'$ too. This fact makes
reusing D-sequents \ti{non-trivial} because it depends on the context.

%
%

%
%
\subsection{Clarifying what D-sequent reusing means}
A few definitions below clarify what safe reusing of a D-sequent
means.
%
%
\begin{definition}
\label{def:dseq_holds}  
We will say that a D-sequent \Oods{q}{U}{C} \tb{holds} for a formula
\prob{X}{F} if \cof{C}{q} is redundant in \prob{X}{\cof{F}{q}} (\ie
$C$ is redundant in \prob{X}{F} in subspace \pnt{q}).
\end{definition}
%
%
\begin{definition}
\label{def:consist}
We will call assignments \pent{q}{'} and \pent{q}{''} \tb{consistent}
if every variable of $\V{\pent{q}{'}} \cap \V{\pent{q}{''}}$ has
identical values in \pent{q}{'} and \pent{q}{''}.
\end{definition}
%
%
\begin{definition}
  \label{def:main}
Let $\oDs{\ppnt{q}{0}}{U_0}{C_0},\dots,\oDs{\ppnt{q}{i}}{U_i}{C_i}$ be
a set of D-sequents with consistent conditionals that hold for
\prob{X}{F} \ti{individually}. Let \pent{q}{'} be an \ti{arbitrary}
assignment where $\pent{q}{'} \supseteq \ppnt{q}{0},\dots, \pent{q}{'}
\supseteq \ppnt{q}{i}$.  The reuse of the D-sequents above means
\ti{removing} the clauses \s{C_0,\dots,C_i} \ti{together} from
\prob{X}{F} in subspace \pent{q}{'} as redundant.  We call such a
reuse \tb{safe} if these clauses are indeed jointly redundant in
subspace \pent{q}{'}.
\end{definition}

\section{Reusing Single D-sequent And Scale-Down Property}
\label{sec:sngl_dseq}
According to Definition~\ref{def:main}, to be safely reused, a
D-sequent \Oods{q}{U}{C} must have at least the property that, for
every $\pent{q}{'} \supseteq \pnt{q}$, the D-sequent
\oDs{\pent{q}{'}}{U}{C} holds too. (From now on when we say a
D-sequent we mean an \tb{extended} D-sequent.) That is, if $C$ is
redundant in subspace \pnt{q} it should be redundant in any subspace
\ti{contained} in subspace \pnt{q}. If so, we will say that the
D-sequent above has the \tb{scale-down} property. In general,
redundancy of a clause does not scale down (see
Example~\ref{exmp:scale_down}). Previously, to get around this
problem, we used a convoluted and counter-intuitive definition of
clause redundancy in a subspace~\cite{fmsd14}. In this section, we
describe a much nicer and simpler solution based on the notion of
\ti{constructive} D-sequents. The latter are exactly the D-sequents
produced by the existing algorithms.  We show that constructive
D-sequents either have the scale-down property
(Proposition~\ref{prop:main_sngl}) or violating this property does not
affect their reusability (Remark~\ref{rem:scale_down}).

%
%
\begin{example}
\label{exmp:scale_down}
Let \prob{X}{F(X)} be a formula where $F= C_0\wedge C_1 \wedge C_2$
and $C_0 = x_0 \vee x_1$, $C_1=x_0 \vee \overline{x}_1$,
$C_2=\overline{x}_0 \vee x_1$ and $X=\s{x_0,x_1}$. Since $F$ is
satisfiable, $C_0$ is redundant in \prob{X}{F} in the entire space
(because \prob{X}{F} = \prob{X}{F \setminus \s{C_0}} = 1). Let us show
that this redundancy \ti{does not scale down}. Consider the subspace
$\pent{q}{'} = (x_0=0)$. In this subspace, $(C_0)_{\vec{q}\,'} = x_1$,
$(C_1)_{\vec{q}\,'} = \overline{x}_1$, $(C_2)_{\vec{q}\,'} \equiv
1$. So, the formula $F$ is unsatisfiable in subspace \pent{q}{'}.
Note that removing $C_0$ makes $F$ satisfiable in subspace
\pent{q}{'}.  So, $C_0$ is \ti{not} redundant in subspace \pent{q}{'}
being redundant in the entire space.
\end{example}

%
%
\begin{definition}
A D-sequent is called \tb{constructive} if
\begin{itemize}
\item it is an atomic D-sequent or
\item it is obtained by resolving two constructive D-sequents
\end{itemize}
\end{definition}

%
%
\begin{proposition}
  \label{prop:main_sngl}
Let \Oods{q}{U}{C} be a constructive D-sequent stating the redundancy
of a clause $C$ in \prob{X}{F} in subspace \pnt{q}. Assume that its
derivation does not involve D-sequents stating redundancy of clauses
other than $C$. Then this D-sequent has the scale-down property.
\end{proposition}
%
%
\begin{remark}
  \label{rem:scale_down}
Let $D$ denote the D-sequent \Oods{q}{U}{C} from
Proposition~\ref{prop:main_sngl}.  In this proposition, we assume that
when deriving $D$ one did not use redundancy of clauses different from
$C$. This assumption essentially means that no clause of $U$ has been
removed as redundant in subspace \pnt{q} by the time $D$ is
derived. If this assumption is broken, one cannot guarantee that $D$
has the scale-down property. However, as we show in the next section
one can still safely reuse $D$ even if some clauses of $U$ are removed
as redundant in subspace \pnt{q}. The only exception (\ie the case
when $D$ \ti{cannot} be reused) is given in
Proposition~\ref{prop:main_mult}.
\end{remark}
%
%
\begin{example}
Let us show how one can express the redundancy of clause $C_0$ from
Example~\ref{exmp:scale_down} by a constructive D-sequent. One can
build such a D-sequent by branching on $x_0$. In the branch $x_0=0$,
one has to add to $F$ the conflict clause $K\!=\!x_0$ to \ti{make}
$C_0$ redundant in subspace $x_0=0$. Then the atomic D-sequent
\oDs{\pent{q}{'}}{U'}{C_0} holds for \prob{X}{F \wedge K} where
$\pent{q}{'} = (x_0=0)$ and $U' = \s{K}$. In the branch $x_1=1$, the
atomic D-sequent \oDs{\pent{q}{''}}{U''}{C_0} holds since $C_0$ is
satisfied by $\pent{q}{''} = (x_0=1)$. Here $U'' = \emptyset$. By
resolving these D-sequents on $x_0$ one derives a constructive
D-sequent \Oods{q}{U}{C_0} for the formula \prob{X}{F \wedge K} where
$\pnt{q} = \emptyset$ and $U = \s{K}$. This D-sequent expresses the
global redundancy of $C$ in \prob{X}{F \wedge K} (instead of
\prob{X}{F}) and has the scale-down property.
\end{example}

\section{Reusing Multiple D-sequents}
\label{sec:mltpl_dseqs}
In \cite{qe_learn} some severe limitations were imposed on reusing an
extended D-sequent. This was done to avoid circular reasoning when
multiple D-sequents are reused \ti{jointly}.  In this section, we show
that one can \ti{lift those limitations} without risking to produce
the wrong result. Arguably, this will make reusing D-sequents
dramatically more effective.

Let $C$ be a clause of a formula \prob{X}{F}. Let $D$ denote an
D-sequent \Oods{q}{U}{C}.  In~\cite{qe_learn}, we showed that $D$ can
be safely reused only if all clauses of $U$ are present in the current
formula \ie none of the clauses of $U$ is removed from \prob{X}{F} in
subspace \pnt{q} as satisfied or redundant. This is very
restrictive. Indeed, a clause of $U$ can be in two states: present/not
present in the formula. So, the total number of possible states of $U$
is $2^{|U|}$. The restriction above means that $D$ can be reused only
in \tb{one} of $2^{|U|}$ cases. The proposition below \tb{lifts this
  restriction}.

%
%
\begin{proposition}
  \label{prop:main_mult}
Given a formula \prob{X}{F} and a derived D-sequent \Oods{q}{U}{C},
the latter can be safely reused in subspace $\pent{q}{^*} \supseteq
\pnt{q}$ with one exception. Namely, this D-sequent cannot be reused
if another D-sequent \oDs{\pent{q}{'}}{U'}{C'} was applied earlier
where $C' \in U$ and $C \in U'$ (and $\pent{q}{^*} \supseteq
\pent{q}{'}$).
\end{proposition}

Proposition~\ref{prop:main_mult} dramatically boosts the applicability
of a D-sequent \Oods{q}{U}{C}. It allows to reuse this D-sequent
\ti{regardless} of whether clauses of $U$ are present in subspace
\pent{q}{^*}. The only \ti{exception} to such reusing is that a
removed clause $C' \in U$ is proved redundant using $C$ itself. (Then
reusing \Oods{q}{U}{C} leads to circular reasoning.)
%
%
\begin{example}
Let \prob{X}{F} be a formula where $F=C_0 \wedge C_1 \wedge C_2 \wedge
\dots$ and $C_0=x_0\vee x_1 \vee x_2$, $C_1=x_1 \vee x_2 \vee x_3$,
$C_2 = \overline{x}_0 \vee \overline{x_2}$, and $x_i \in X$.  Since
$C_0$ implies $C_1$ in subspace $\pnt{q}=(x_0\!=\!0)$, the D-sequent
\Oods{q}{U}{C_1} holds where\linebreak $U = \s{C_0}$. It states that
$C_0$ was used to prove redundancy of $C_1$ in subspace
\pnt{q}. Assume that $C_2$ is the only clause of $F$ with the literal
$\overline{x}_0$. Then $F$ has no clauses resolvable with $C_0$ on
$x_0$ and hence $C_0$ is blocked at $x_0$. So, the D-sequent
\oDs{\pent{q}{'}}{U'}{C_0} holds where $\pent{q}{'} = \emptyset$,
$U'=\emptyset$ and $C_0$ can be removed from $F$ as redundant.
According to Proposition~\ref{prop:main_mult}, the D-sequent
\Oods{q}{U}{C_1} still can be \ti{safely reused} even though the
clause $C_0$ of $U$ is removed from the formula.
\end{example}

\section{Some Background}
\label{sec:bckgr}
Information on QE in propositional logic can be found
in~\cite{blocking_clause,cav09,cav11,cmu,nik2,cadet_qe}.  QE by
redundancy based reasoning is presented in~\cite{fmcad12,fmcad13}.
One of the merits of such reasoning is that it allows to introduce
\ti{partial} QE.  A description of PQE algorithms and their sources
can be found in~\cite{hvc-14,cav23,cert_tech_rep,ds_pqe,eg_pqe_plus}.

Removal of redundancies is used in pre-processing
procedures~\cite{prepr,inproc,prop_red}. Such procedures typically
look for shallow redundancies that can be easily identified (e.g. they
search for clauses that are trivially blocked). The objective here is
to \ti{minimize/simplify} the formula at hand. A PQE solver can also
be viewed as a tool for identifying redundancies. In particular, in
some applications (see \eg~\cite{cav23}) it is simply used to check if
a subset of clauses $G$ is redundant in \prob{X}{F}. This version of
the problem is called a \ti{decision} PQE problem. However, the goal
here is to solve the problem rather than optimize \prob{X}{F}. So, the
PQE solver keeps proving redundancy of $G$ even if this requires
exploring a deep search tree. The general case of the PQE problem
where one takes $G$ out of the scope of quantifies is even further
away from optimization of \prob{X}{F}. Here one needs to \ti{make} $G$
redundant by adding a formula $H(Y)$ whose size can be much larger
than that of $G$.

In~\cite{cert_tech_rep} we introduced the notion of a certificate
clause that can be viewed as an advanced form of a D-sequent.  The
idea here is to prove a clause $C$ redundant in subspace \pnt{q} by
deriving a clause $K$ that \ti{implies} $C$ in subspace \pnt{q}.  The
clause $K$ is called a certificate. The advantage of a certificate
clause over a D-sequent is that the former can be added to \prob{X}{F}
(because $\prob{X}{F} \equiv \prob{X}{F \wedge K}$).

The introduction of certificate clauses addresses the problem of
reusing the information learned by a PQE solver. The presence of $K$
in the formula makes the redundancy of the clause $C$ above in
subspace \pnt{q} obvious. However the machinery of certificates has
the following flaw. The certificate $K$ above can be represented as
$C' \vee C''$ where $C'$ is falsified by \pnt{q} and $C''$ consists of
literals of $C$. If $C''$ contains all the literals of $C$, then
adding $K$ to \prob{X}{F} does not make sense (because $K$ is implied
by $C$).  If one uses $K$ as a proof of redundancy in subspace \pnt{q}
\ti{without} adding it to \prob{X}{F}, one runs into the same problem
as with regular D-sequents. Namely, indiscriminate reusing of the
certificate $K$ can lead to circular reasoning.

\section{Conclusions}
\label{sec:concl}
We consider the problem of taking a subset of clauses $G$ out of the
scope of quantifiers in a propositional formula \prob{X}{F}. We refer
to this problem as partial quantifier elimination (PQE). We solve PQE
using the machinery of D-sequents where a D-sequent is a record
stating that a clause $C$ is redundant in \prob{X}{F} in subspace
\pnt{q}.  We use this machinery because in contrast to a SAT algorithm
enumerating only the subspaces where the formula is unsatisfiable, a
PQE solver has to also enumerate those where the formula is
\ti{satisfiable}. The main flaw of the current PQE solvers is that
they do not reuse D-sequents.  Reusing D-sequents is not trivial
because redundancy of a clause is a structural rather than semantic
property. We show that by using an extended version of D-sequents one
can safely reuse them. The next natural step is show the viability of
our approach \ti{experimentally}.

\bibliographystyle{IEEEtran}
\bibliography{short_sat,local,l1ocal_hvc}
\appendix
\noindent{\large \tb{Appendix}}
\section{Proofs Of Propositions}
 \setcounter{proposition}{0}
 \label{app:proofs}

%
%
\subsection{Propositions of Section~\ref{sec:basic}}
%
%
\begin{proposition}
Let a clause $C$ be blocked in a formula $F(X,Y)$ with respect to a
variable $x \in X$.  Then $C$ is redundant in \prob{X}{F}
i.e., \prob{X}{F \setminus \s{C}} $\equiv$ \prob{X}{F}.
\end{proposition}
\begin{proof}
It was shown in~\cite{blocked_clause} that adding a clause $B(X)$
blocked in $G(X)$ to the formula \prob{X}{G} does not change the value
of this formula.  This entails that removing a clause $B(X)$ blocked
in $G(X)$ does not change the value of \prob{X}{G} either. So, $B$ is
redundant in \prob{X}{G}.

Now, let us return to the formula \prob{X}{F(X,Y)}. Let \pnt{y} be a
full assignment to $Y$. Then the clause $C$ of the proposition at hand
is either satisfied by \pnt{y} or \cof{C}{y} is blocked in \cof{F}{y}
with respect to $x$. (The latter follows from the definition of a
blocked clause.) In either case, \cof{C}{y} is redundant in
\prob{X}{\cof{F}{y}}. Since this redundancy holds in every subspace
\pnt{y}, the clause $C$ is redundant in \prob{X}{F}.
\end{proof}

%
%
\subsection{Propositions of Section~\ref{sec:pqe_alg}}
%
%
\begin{proposition}
  Formula $H(Y)$ is a solution to the PQE problem of taking 
  $G$ out of \prob{X}{F(X,Y)} (\ie $ \prob{X}{F} \equiv H \wedge
  \prob{X}{F \setminus G}$) iff
  \begin{enumerate}
  \item $F \imp H$ and
  \item $H \wedge
    \prob{X}{F} \equiv H \wedge \prob{X}{F \setminus G}$
  \end{enumerate}
\end{proposition}
\begin{proof}
\noindent\tb{The if part.} Assume that conditions 1, 2 hold. Let us
show that $\prob{X}{F} \equiv H \wedge \prob{X}{F \setminus
  G}$. Assume the contrary \ie there is a full assignment \pnt{y} to
$Y$ such that $\prob{X}{F} \neq H \wedge \prob{X}{F \setminus G}$
in subspace \pnt{y}.

There are two cases to consider here. First, assume that $F$ is
satisfiable and $H \wedge (F \setminus G)$ is unsatisfiable in
subspace \pnt{y}. Then there is an assignment (\pnt{x},\pnt{y})
satisfying $F$ (and hence satisfying $F \setminus G$).  This means
that ($\pnt{x},\pnt{y}$) falsifies $H$ and hence $F$ does not imply
$H$.  So, we have a contradiction. Second, assume that $F$ is
unsatisfiable and $H \wedge (F \setminus G)$ is satisfiable in
subspace \pnt{y}. Then $H \wedge F$ is unsatisfiable too. So,
condition 2 does not hold and we have a contradiction.

\noindent\tb{The only if part.} Assume that $ \prob{X}{F} \equiv H
\wedge \prob{X}{F \setminus G}$. Let us show that conditions 1 and 2
hold.  Assume that condition 1 fails \ie $F \not\imp H$. Then there is
an assignment (\pnt{x},\pnt{y}) satisfying $F$ and falsifying
$H$. This means that $ \prob{X}{F} \neq H \wedge \prob{X}{F \setminus
  G}$ in subspace \pnt{y} and we have a contradiction. To prove that
condition 2 holds, one can simply conjoin both sides of the equality
$ \prob{X}{F} \equiv H \wedge \prob{X}{F \setminus G}$ with $H$.
\end{proof}

%
%
\subsection{Propositions of Section~\ref{sec:sngl_dseq}}
Lemmas~\ref{lem:align}-\ref{lem:res} are used in the proof
of Proposition~\ref{prop:main_sngl}.
%
%
\begin{lemma}
\label{lem:align}
Let \Oods{q}{U}{C} be a D-sequent that holds for a formula
\,\prob{X}{F(X,Y)} where $C$ is a quantified clause of $F$. Then this
D-sequent also holds for any formula \prob{X}{F'} obtained from
\prob{X}{F} by adding clauses implied by $F$.
\end{lemma}
\begin{proof}
The fact that \Oods{q}{U}{C} holds means that \cof{C}{q} is redundant
in \prob{X}{\cof{F}{q}}. Denote \cof{C}{q} as $C^*$ and \cof{F}{q} as
$F^*$. Redundancy of $C^*$ in \prob{X}{F^*} means that there is no
full assignment \pnt{y} to $Y$ such that \cof{F^*}{y} is unsatisfiable
and \cof{(F^* \setminus \s{C^*})}{y} is satisfiable. In other words,
the fact that \cof{F^*}{y} is unsatisfiable implies that \cof{(F^*
  \setminus \s{C^*})}{y} is unsatisfiable too. The same is true if
$F^*$ (\ie \cof{F}{q}) is replaced with \cof{F'}{q}.
\end{proof}

Lemma~\ref{lem:align} allows to ``align'' D-sequents derived at
different times so that they can be resolved. Suppose a D-sequent
\Oods{q}{U}{C} was derived for \prob{X}{F}. Suppose another D-sequent
\oDs{\pent{q}{'}}{U'}{C} was derived later for a formula \prob{X}{F'}
where $F'$ was obtained from $F$ by adding clauses implied by
$F$. Then according to Lemma~\ref{lem:align}, the D-sequent
\Oods{q}{U}{C} holds for \prob{X}{F'} too.  So, by resolving these two
D-sequents one obtains a D-sequent that holds for \prob{X}{F'}.
%
%
\begin{lemma}
\label{lem:sat}
Let $C$ be a quantified clause of\, \prob{X}{F}. Let $C$ be satisfied
in subspace \pnt{q} and thus redundant there.  Let \Oods{q}{U}{C} be
the D-sequent stating this redundancy of $C$ where $U = \emptyset$.
Then this D-sequent has the scale-down property.
\end{lemma}
\begin{proof}
If \pnt{q} satisfies $C$, then every $\pent{q}{'} \supseteq \pnt{q}$
satisfies $C$ too. So, \oDs{\pent{q}{'}}{U}{C} holds.
\end{proof}

%
%
\begin{lemma}
\label{lem:impl}
Let $C$ be a quantified clause of\, \prob{X}{F}. Let $C$ be implied by
another clause $B \in F$ in subspace \pnt{q} so, the D-sequent
\Oods{q}{U}{C} holds where $U = \s{B}$. This D-sequent has the
scale-down property.
\end{lemma}
\begin{proof}
If $B$ implies $C$ in subspace \pnt{q}, it also implies $C$ in every
subspace $\pent{q}{'} \supseteq \pnt{q}$. So, \oDs{\pent{q}{'}}{U}{C}
holds.
\end{proof}

%
%
\begin{lemma}
\label{lem:block}
Let $C$ be a quantified clause of \prob{X}{F}. Let every clause of $F$
resolvable with $C$ on $x \in X$ be satisfied by an assignment
\pnt{q}.  So, $C$ is blocked at $x$ in subspace \pnt{q} and hence
redundant in this subspace.  Then the D-sequent \Oods{q}{U}{C}, $U =
\emptyset$, stating this redundancy has the scale-down property.
\end{lemma}
\begin{proof}
Let \pent{q}{'} be a superset of \pnt{q}. Then we have two
possibilities.  First, \pent{q}{'} satisfies $C$. Then the latter is
redundant in subspace \pent{q}{'}. Second, \pent{q}{'} does not
satisfy $C$. Then the fact that \pnt{q} satisfies all clauses of $F$
resolvable with $C$ on $x$ implies that the same is true for
\pent{q}{'}. Hence $C$ is blocked in subspace \pent{q}{'} at $x$.  In
either case, $C$ is redundant in subspace \pent{q}{'} and so the
D-sequent \oDs{\pent{q}{'}}{U}{C} holds.
\end{proof}

Note that Lemma~\ref{lem:block} holds only for a \ti{special} case of
being blocked.  In general, when $C$ is blocked in subspace \pnt{q},
some clauses resolvable with $C$ on $x$ are \ti{proved} redundant in
subspace \pnt{q} rather than satisfied by \pnt{q}. Then the
construction set $U$ can be non-empty.
%
%
\begin{lemma}
\label{lem:res}
Let $C$ be a clause of a formula $F$.  Let \Oods{r}{U^*}{C} and
\Oods{s}{U^{**}}{C} be D-sequents that hold for \prob{X}{F} and that
can be resolved on a variable $v$.  Let these D-sequents have the
scale-down property. Then the D-sequent \Oods{q}{U}{C}, $U=U^* \cup
U^{**}$ obtained by resolving these D-sequents on $v$ has the
scale-down property too.
\end{lemma}
\begin{proof}
Recall that by the definition of resolution of D-sequents, \pnt{q}
equals $\pnt{r} \cup \pnt{s}$ minus assignments to $v$.  Assume, for
the sake of clarity, that $v=0$ in \pnt{r} and $v=1$ in \pnt{s}. Let
$\pent{q}{'} \supseteq \pnt{q}$ hold. One needs to consider the
following two possibilities. First, assume that \pent{q}{'} assigns a
value to $v$. If $v=0$, then $\pent{q}{'} \supseteq \pnt{r}$ holds and
$C$ is redundant in subspace \pent{q}{'} (so, \oDs{\pent{q}{'}}{U}{C}
holds). If $v=1$, then $\pent{q}{'} \supseteq \pnt{s}$ holds and $C$
is redundant in subspace \pent{q}{'} as well. Second, assume that
\pent{q}{'} \ti{does not} assign any value to $v$. Let us split the
subspace \pent{q}{'} into subspaces $\vec{q}\,'_0$ and $\vec{q}\,'_1$
where $\vec{q}\,'_0 = \pent{q}{'} \cup \s{(v=0)}$ and $\vec{q}\,'_1 =
\pent{q}{'} \cup \s{(v=1)}$. Since $\vec{q}\,'_0 \supseteq \pnt{r}$,
the clause $C$ is redundant in subspace $\vec{q}\,'_0$. Since
$\vec{q}\,'_1 \supseteq \pnt{s}$, $C$ is redundant in subspace
$\vec{q}\,'_1$ too.  So, $C$ is redundant in subspace \pent{q}{'}.
\end{proof}

%
%
\begin{proposition}
Let \Oods{q}{U}{C} be a constructive D-sequent stating the redundancy
of a clause $C$ in \prob{X}{F} in subspace \pnt{q}. Assume that its
derivation does not involve D-sequents stating redundancy of clauses
other than $C$. Then this D-sequent has the scale-down property.
\end{proposition}
%
%
\begin{proof}
Assume, for the sake of clarity, that D-sequents are generated by
\Dpqe.  Let \aps{D} denote the set of D-sequents involved in
generating the D-sequent at hand. By the assumption of our
proposition, every D-sequent of \aps{D} specifies redundancy of the
clause $C$. We prove this proposition by induction on the order in
which the D-sequents of \aps{D} are derived by \Dpqe.  Denote by $D_n$
the D-sequent with the order number $n$.  First, let us prove the base
case \ie this proposition holds for the first D-sequent $D_1$. Let
$D_1$ be equal to \oDs{\ppnt{q}{1}}{U_1}{C}. The clause $C$ is either
satisfied, implied by another clause or blocked in subspace
\ppnt{q}{1}.  Then according to Lemmas~\ref{lem:sat},\ref{lem:impl}
and~\ref{lem:block}, $D_1$ has the scale-down property.

Now, we prove that the D-sequent $D_n$ has the scale-down property
assuming that this property holds for the D-sequents
$D_1,\dots,D_{n-1}$. Let $D_n$ be equal to \oDs{\ppnt{q}{n}}{U_n}{C}
and $\pent{q}{'} \supseteq \ppnt{q}{n}$. Assume that $C$ is redundant
in subspace \ppnt{q}{n} because it is satisfied there or implied by an
existing clause of $F$. Then, according to
Lemmas~\ref{lem:sat},\ref{lem:impl}, the D-sequent $D_n$ has the
scale-down property.  Now, assume that $C$ is blocked in subspace
\ppnt{q}{n} at a variable $x \in X$. Since D-sequents of clauses other
than $C$ are not used in derivation of the target D-sequent
\Oods{q}{U}{C}, the construction set $U_n$ is empty. (If $U_n \neq
\emptyset$, at least one clause resolvable with $C$ on $x$ is not
present in the formula because it was proved redundant in subspace
\pnt{q} rather than satisfied by \pnt{q}.) Then, one can apply
Lemma~\ref{lem:block} to claim that $D_n$ scales down. Finally, assume
that $D_n$ is obtained by resolving D-sequents $D_i$ and $D_j$ where
$i,j < n$. Then according to Lemma~\ref{lem:res}, the D-sequent $D_n$
has the scale-down property too.
\end{proof}

%
%
\subsection{Propositions of Section~\ref{sec:mltpl_dseqs}}
Again, we assume, for the sake of clarity, that D-sequents are
generated by \Dpqe. Before proving Proposition~\ref{prop:main_mult},
we need to give a few definitions.  Let $D$ denote a D-sequent
\Oods{q}{U}{C}. We will say that $D$ is \tb{active} if \Dpqe entered a
subspace \pent{q}{'} where $\pent{q}{'} \supseteq \pnt{q}$ and
temporarily removed the clause $C$ as redundant. We will say that $D$
is \ti{reused} if it is an old D-sequent that was already active at
least once and then became inactive when \Dpqe moved to another
subspace. We will say that $D$ is \ti{used} if we do not care whether
$D$ is a new D-sequent applied for the first time or it is an old
D-sequent being reused.

%
%
\begin{proposition}
Given a formula \prob{X}{F} and a derived D-sequent \Oods{q}{U}{C},
the latter can be safely reused in subspace $\pent{q}{^*} \supseteq
\pnt{q}$ with one exception. Namely, this D-sequent cannot be reused
if another D-sequent \oDs{\pent{q}{'}}{U'}{C'} was applied earlier
where $C' \in U$ and $C \in U'$ (and $\pent{q}{^*} \supseteq
\pent{q}{'}$).
\end{proposition}
\begin{proof}
We will prove this proposition by induction on the number of active
D-sequents. Let $D_n$ denote the $n$-th active D-sequent used by
\Dpqe. Let us prove the base case that the D-sequent $D_1$ can be
safely used. Let $D_1$ be equal to \oDs{\ppnt{q}{1}}{U_1}{C_1}.  Since
$D_1$ is the only active D-sequent so far no other D-sequent was used
in derivation of $D_1$.  Then from Proposition~\ref{prop:main_sngl} it
follows that $D_1$ holds in subspace \pent{q}{^*}.

Now, we prove the induction step. Namely, we show that a D-sequent
$D_n$ can be safely used under the assumption that the D-sequents
$D_1,\dots,D_{n-1}$ have already been correctly applied. Let $D_n$ be
equal to \oDs{\ppnt{q}{n}}{U_n}{C_n}. If every clause of the set $U_n$
is present \prob{X}{F} or satisfied by the current assignment
\pent{q}{^*}, then from Proposition~\ref{prop:main_sngl} it follows
that $D_n$ can be safely used. Now assume that some clauses of $U_n$
have been temporarily removed from \prob{X}{F} as redundant. Let $C_i$
be one of such clauses where $i < n$.  Let $D_i$ denote the D-sequent
that was used by \Dpqe to remove $C_i$ from \prob{X}{F} in subspace
\pent{q}{^*}. Let $D_i$ be equal to \oDs{\ppnt{q}{i}}{U_i}{C_i}.

Let us temporarily add $C_i$ (and the other clauses of $U_n$ removed
as redundant) back to \prob{X}{F} to obtain a formula
\prob{X}{F^*}. Adding these clauses to \prob{X}{F} is a safe operation
because they are \ti{implied} by the original formula $F$. (\Dpqe adds
only conflict clauses and they are implied by the original formula
$F$.) Since every clause of $U_n$ is either present in \prob{X}{F^*}
or satisfied, one can safely apply the D-sequent $D_n$ to remove $C_n$
from \prob{X}{F^*}. Our final step is to remove $C_i$ (and the other
clauses we temporarily added to the current formula) again. This can
be done for the following reasons.  First, by the induction
hypothesis, the D-sequent $D_i$ can be safely used for \prob{X}{F} in
subspace \pent{q}{^*}. Second, the proposition at hand implies that
the set $U_i$ of $D_i$ does not contain $C_n$. So $D_i$ holds for
\prob{X}{F \setminus \s{C_n}} too (because $C_n$ was not used in
proving $C_i$ redundant). Third, using Lemma~\ref{lem:align} one can
claim that $D_i$ holds for \prob{X}{F^* \setminus \s{C_n}} and so can
be safely used.  After removing $C_i$ and the other clauses that were
temporarily added, one recovers the current formula \prob{X}{F} but
without the clause $C_n$. So, the D-sequent $D_n$ has been safely
applied to \prob{X}{F} in subspace \pent{q}{^*}.
\end{proof}

\section{Pseudocode of \Dpqe}
\label{app:pseudo_code}

The pseudocode of \Dpqe is shown in Figure~\ref{fig:ds-pqe}. (A more
detailed description of \Dpqe can be found in~\cite{hvc-14}.) We
assume here that \Dpqe derives ``regular'' D-sequents introduced
in~\cite{fmcad13} and does not reuse them. Since \Dpqe recursively
calls itself, it accepts four parameters: a formula \prob{X}{F}, the
subset $G$ of $F$ to take out, an assignment \pnt{q} and a set of
D-sequents \ti{Ds}. This call of \Dpqe solves the PQE problem of
taking $G$ out of \prob{X}{F} in subspace \pnt{q}.  The set
\ti{Ds} contains D-sequents of the clauses of $F$ already proved
redundant in subspace \pnt{q} (in earlier calls of \Dpqe). In the
first call of \Dpqe both \pnt{q} and \ti{Ds} are empty.  \Dpqe returns
the current formula $F$ and the set \ti{Ds} of D-sequents stating the
redundancy of each clause of $G$ in subspace \pnt{q}, \ie $|\mi{Ds}| =
|G|$.  The solution $H(Y)$ of the PQE problem produced by \Dpqe
consists of the unquantified clauses of the \ti{final}
formula \prob{X}{F} added to the \ti{initial} formula \prob{X}{F}. The
initial (respectively final) formula \prob{X}{F} is passed over in the
first call of \Dpqe (respectively returned by the first call). 
%
%
\setlength{\intextsep}{4pt}
\setlength{\textfloatsep}{10pt}
\begin{wrapfigure}{lh}{1.2in}
\footnotesize
\begin{tabbing}
aaa\=bb\= cc\= ddddddd\= \kill
// $\xi$ denotes the PQE problem \\
// \ie (\prob{X}{F},$G$) \\
// \\
$\dpqe(\xi,\pnt{q},\mi{Ds})$\{\\
\tb{\scriptsize{1}}\> $\pent{q}{^*} := \mi{bcp}(F,G,\pnt{q})$ \}\\
\tb{\scriptsize{2}}\> if $(\neg\mi{done}(\xi,\pent{q}{^*},\mi{Ds}))$ \\
\tb{\scriptsize{3}}\Tt   goto $\mi{Branch}$ \\
\tb{\scriptsize{4}}\> if $(\mi{confl}(F,\pent{q}{^*}))$  \{ \\
\tb{\scriptsize{5}}\Tt   $K := \mi{cnfl\_cls}(F,\pnt{q},\pent{q}{^*})$\\
\tb{\scriptsize{6}}\Tt   $F := F \cup \s{K}$  \\
\tb{\scriptsize{7}}\Tt $\mi{Ds} := \mi{dseqs}_1(\xi,\pnt{q},K,\mi{Ds})$\}\\
\tb{\scriptsize{8}}\> else \\
\tb{\scriptsize{9}}\Tt $\mi{Ds}\!:=\!\mi{dseqs}_2(\xi,\!\pnt{q},\!\pent{q}{^*},\!\mi{Ds})$ \\
\tb{\scriptsize{10}}\> return($\xi,\mi{Ds}$)\\
\tb{\scriptsize{11}} $\mi{Branch}:$~~~//$\text{-\,-\,-\,-\,-\,-\,-\,-\,-\,-}$\\
\tb{\scriptsize{12}} $G := \mi{add\_trgs}(F,G,\pent{q}{^*})$ \\
\tb{\scriptsize{13}}\> $v := \mi{pick\_var}(\pent{q}{^*},X,Y)$   \\
\tb{\scriptsize{14}}\> $\pent{q}{'} := \pent{q}{^*} \cup \s{v=0}$  \\
\tb{\scriptsize{15}}\> $(F,\mi{Ds}')\!:=\!\dpqe(\xi,\pent{q}{'},\mi{Ds})$\\
\tb{\scriptsize{16}}\> if $(\mi{done}(\xi,\pent{q}{^*},\mi{Ds}'))$ \\
\tb{\scriptsize{17}}\Tt  \{$\mi{Ds} := \mi{Ds}'$; goto $\mi{Finish}$\} \\
\tb{\scriptsize{18}}\> $\pent{q}{''} := \pent{q}{^*} \cup \s{v=1}$  \\
\tb{\scriptsize{19}}\> $(F,\mi{Ds}'')\!:=\!\dpqe(\xi,\pent{q}{''},\mi{Ds})$\\
\tb{\scriptsize{20}}\>  $\mi{Ds} := \mi{res\_dseqs}(\mi{Ds}',\mi{Ds}'',v)$ \\
\tb{\scriptsize{21}} $\mi{Finish}:$  ~~~//$\text{-\,-\,-\,-\,-\,-\,-\,-\,-\,-}$\\
\tb{\scriptsize{22}}\>  $G := \mi{rmv\_trgs}(G,\pnt{q},\pent{q}{^*})$ \\
\tb{\scriptsize{23}}\>  $\mi{Ds} := \mi{shorten}(F,\mi{Ds},\pnt{q},\pent{q}{^*})$ \\
\tb{\scriptsize{24}}\>  return($\xi,\mi{Ds}$)\}\\
\end{tabbing} 
\vspace{-17pt}
\caption{\Dpqe}
\vspace{-10pt}
\label{fig:ds-pqe}
\end{wrapfigure}

%
\Dpqe starts by checking if $F$ has unit clauses in subspace \pnt{q}
(line 1, of Fig.~\ref{fig:ds-pqe}).  If so, it runs Boolean Constraint
Propagation (BCP) extending the assignment \pnt{q} to \pent{q}{^*}.
Assume that there is at least one clause of $G$, for which no
\ti{atomic} D-sequent can be derived in subspace \pent{q}{^*} (see
Definition~\ref{def:atomic}). Then \Dpqe goes to the \ti{branching}
part of the algorithm (lines 2-3). Otherwise, \Dpqe derives D-sequents
for the clauses of $G$ that were not proved redundant yet in subspace
\pnt{q} and terminates the current call of \Dpqe (lines 4-10). Namely,
if a conflict occurs in subspace \pent{q}{^*}, \Dpqe derives a
conflict clause $K$ that is falsified by \pnt{q} and adds it to $F$
(lines 5-7). Then it constructs an atomic D-sequent \oods{r}{C} for
each target clause $C$ that is not proved redundant yet in subspace
\pnt{q}. Here \pnt{r} is the smallest subset of \pnt{q} that falsifies
$K$. This D-sequent states that $C$ is redundant in any subspace where
$K$ is falsified. If no conflict occurs, each remaining clause $C$ of
$G$ is proved redundant in subspace \pnt{q} by showing that it is
satisfied, implied by an existing clause or blocked. Then an atomic
D-sequent is derived for such a clause $C$ (line 9). Finally, \Dpqe
returns the current formula \prob{X}{F} and the set of
D-sequents \ti{Ds} for the clauses of $G$ (line 10).

The branching part of \Dpqe is shown in lines 12-20. First, \Dpqe
extends the set of target clauses $G$ (line 12). Namely, for each
target clause $C$ that becomes unit during BCP, \Dpqe adds to $G$ the
clauses of $F$ that are resolvable with $C$ on its unassigned
variable. These new clauses are \ti{temporarily} added to $G$. They
are removed from $G$ in the finish part (line 22). Then, \Dpqe picks a
variable $v$ to branch on (line 13).  \Dpqe assigns unquantified
variables (\ie those of $Y$) before quantified (\ie those of $X$). So,
$v$ is in $X$ only if all variables of $Y$ are already assigned. Then
\Dpqe recursively calls itself to explore the branch $\pent{q}{'} :=
\pent{q}{^*} \cup \s{v=0}$ (lines 14-15). If the conditionals of the
D-sequents from the set $\mi{Ds}'$ returned by \Dpqe do not depend on
variable $v$, the branch $\pent{q}{^*} \cup \s{v=1}$ can be
skipped. So, \Dpqe jumps to the finishing part (lines
16-17). Otherwise, \Dpqe explores the branch $\pent{q}{^*} \cup
\s{v=1}$ (lines 18-19).
Finally, \Dpqe resolves D-sequents of both branches whose conditionals
depend on variable $v$ (line 20). Namely, if the D-sequent of a clause
$C \in G$ found in the first branch contains $v=0$ in its conditional,
it is resolved with the D-sequent of $C$ found in the second branch on
variable $v$ (the conditional of this D-sequent contains the
assignment $v=1$).

The finish part of \Dpqe is shown in lines 22-24. In line 22, \Dpqe
removes from $G$ the target clauses added earlier in line 12. Then
\Dpqe shortens the conditionals of the D-sequents of the target
clauses by getting rid of assignments added to \pnt{q} by BCP (line
23). This procedure is similar to the conflict clause generation by a
SAT-solver (where the latter eliminates from the conflict clause the
implied assignments made at the conflict level during BCP). Finally,
\Dpqe returns the current formula \prob{X}{F} and the D-sequents generated for
the clauses of $G$.

\section{The Benefit Of Reusing D-sequents}
\label{app:reuse_dseqs}

%
%
\begin{wraptable}{lh}{2.7in}
\small
\caption{Repeated generation of the same\\D-sequents}
\vspace{-7pt}
\scriptsize
\begin{center}
\begin{tabular}{|c|c|c|c|c|c|c|c|c|} \hline
name &   \multicolumn{2}{c|}{1} & \multicolumn{2}{c|}{2} & \multicolumn{2}{c|}{3} & \multicolumn{2}{c|}{4}   \\  \cline{2-9}
     & all  &   core & all & core & all & core & all & core    \\  \hline
ex1  & 10,919 &  32   &  8,468 & 69 & 8,002 & 55  & 7,390 & 95     \\ \hline
ex2  & 4,038 & 373 & 2,901 & 393  & 2,339 & 285 & 2,204 & 267  \\ \hline
ex3  &2,282  & 657 & 522 & 3 & 448 & 2 & 348 & 1 \\ \hline
ex4 & 562 & 50  & 211 & 19 & 73 & 8 & 72 & 8 \\ \hline
ex5 & 3,320 & 102 &2,066 &14 & 741 & 15& 741 & 7 \\ \hline
ex6 & 808 & 256 & 404 & 131 & 402 & 236  & 232 & 4 \\ \hline
ex7 &9,546 &5 & 6,687  & 10  & 6,497 & 13 & 4,604 & 1\\ \hline 
\end{tabular}                
\end{center}
\vspace{-3pt}
\label{tbl:reuse}
\end{wraptable}

In this appendix, we describe an experiment showing that \Dpqe tends
to repeatedly generate the same D-sequents. This experiment
substantiates the claim of Section~\ref{sec:reuse_pqe} that reusing
D-sequents should boost the performance of PQE solving. In this
experiment we used PQE problems that can be downloaded
from~\cite{my_hwmcc13}.  Each PQE problem is to take a single clause
out of a formula \prob{X}{F} specifying the set of states of a
sequential circuit reachable in $n$ transitions. The circuits used in
those PQE problems were taken from the HWMCC-13 set~\cite{hwmcc13}.

\begin{wraptable}{l}{2.4in}
\caption{Real names of benchmarks}
\setlength\extrarowheight{2pt}
\scriptsize
\begin{tabular}{|p{18pt}|p{135pt}|} \hline
 short & name used in~\cite{my_hwmcc13}  \\ 
 name  &            \\  \hline
 ex1 &6s106.10.unrem.probs.345 \\ \hline
 ex2 &6s110.9.unrem.probs.3221 \\ \hline
 ex3 &6s254.10.unrem.probs.955 \\ \hline
 ex4 &6s255.10.unrem.probs.1535 \\ \hline
 ex5 &6s255.10.unrem.probs.2167 \\ \hline
 ex6 &bob12m05m.3.unrem.probs.5062 \\ \hline
 ex7 &bob9234specmulti.10.unrem.probs.353\\ \hline
\end{tabular}                
\label{tbl:names}
\end{wraptable}

In this experiment, we used PQE problems that were too hard for \Dpqe
to solve.  So, for each problem, we just ran \Dpqe for 10 seconds and
collected data on D-sequents it generated. In this data, we took into
account only non-atomic D-sequents \ie those whose generation involved
at least one resolution on a \ti{decision variable}. The reason for
that is that the generation of non-atomic D-sequents is
computationally harder than that of atomic D-sequents.  So, the
repeated production of the same non-atomic D-sequents is more
costly. Table~\ref{tbl:reuse} provides results for a sample of 7 PQE
problems. The first column of this table shows the short names
assigned to the PQE problems. The names of the problems under which
they are listed in~\cite{my_hwmcc13} are given in
Table~\ref{tbl:names}.

Let \mbox{$G\!=\!\{C\}$} be the original set of clauses to take out of
\prob{X}{F}.  As we mentioned in Section~\ref{sec:pqe_alg}, \Dpqe
temporarily extends $G$.  So, \Dpqe generated D-sequents not only for
the original clause of $G$ but for many other clauses. In
Table~\ref{tbl:reuse}, for each PQE problem, we present results for
the \ti{four} clauses that were present in $G$ for which \Dpqe
generated the largest number of non-atomic D-sequents. So, the columns
2-9 of Table~\ref{tbl:reuse} specify four pairs.
The first column of a pair shows the total number of non-atomic
D-sequents generated for a particular target clause of $G$. The second
column gives the number of the ``core'' D-sequents that were generated
many times. Consider, for example, the first pair of columns of the
instance \ti{ex1}. The first and second columns of this pair give the
numbers 10,919 and 32 respectively. It means that \Dpqe generated
10,919 non-atomic D-sequents but the latter were just repetitions of
the same 32 D-sequents.
The results of Table~\ref{tbl:reuse} show that \Dpqe generated a lot
of identical D-sequents. So reusing learned D-sequents should be quite
helpful.

\section{Unobservability And D-sequents}
\label{app:unobs}
\setlength{\intextsep}{4pt}
\begin{wrapfigure}{lh}{2in}
 \begin{center}
    \includegraphics[width=1.6in]{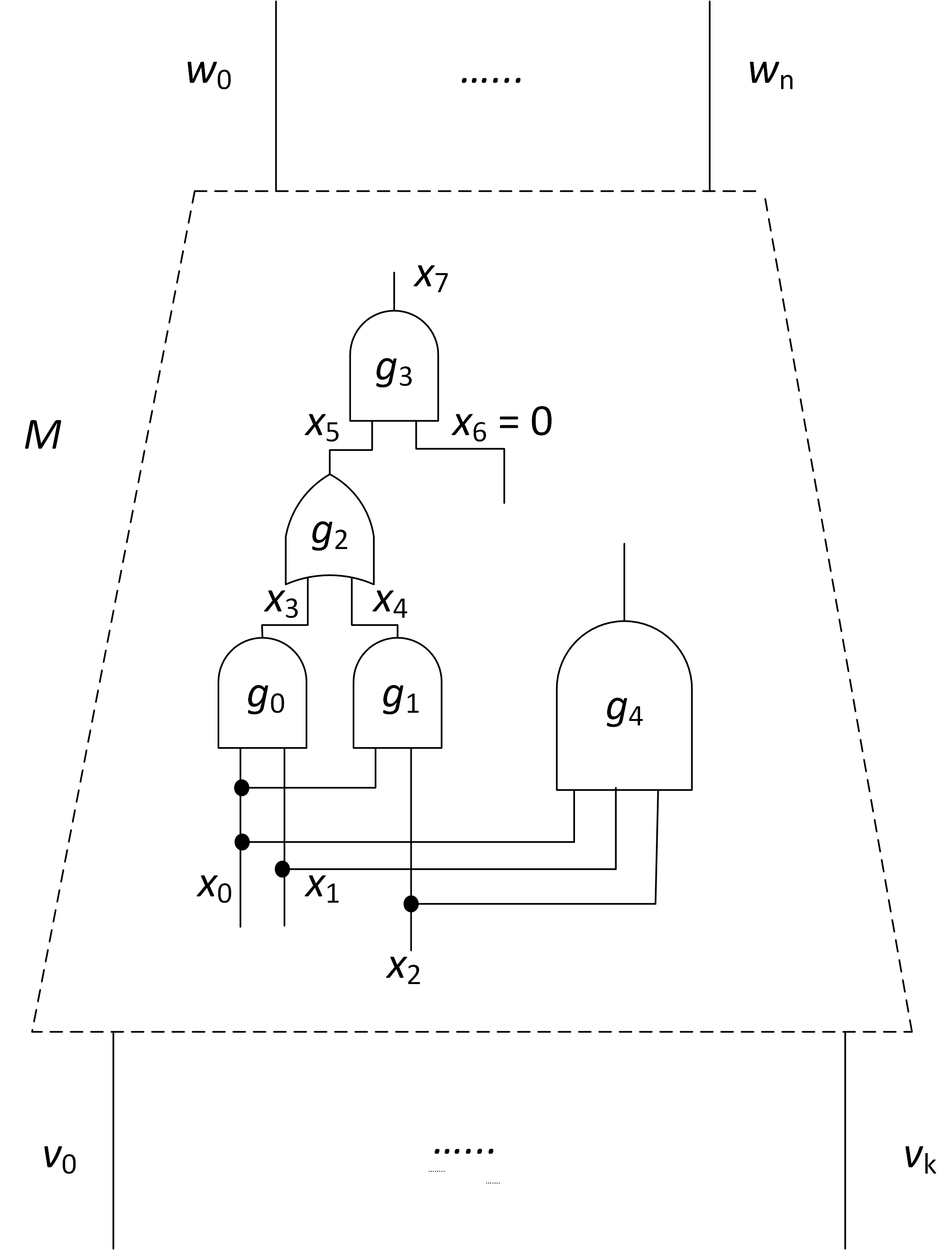}
  \end{center}
\caption{Unobservable gates}
\vspace{2pt}
\label{fig:loc_sat}
\end{wrapfigure}

In this appendix, we illustrate the relation between unobservability
and D-sequents by an example. Consider the fragment of circuit
$M(X,V,W)$ shown in Figure~\ref{fig:loc_sat}. Here $X,V,W$ are sets of
internal, input and output variables respectively. Let $F(X,V,W) =
F_{g_0} \wedge F_{g_1} \wedge F_{g_2} \wedge F_{g_3} \wedge \dots$ be
a formula specifying $M$ where $F_{g_i}$ describes the functionality
of gate $g_i$.  Namely, \bm{F_{g_0}}$ = C_0 \wedge C_1 \wedge C_2$
where $C_0 = \overline{x_0} \vee \overline{x_1} \vee x_3$, $C_1 = x_0
\vee \overline{x_3}$, $C_2 = x_1 \vee \overline{x_3}$, \bm{F_{g_1}}$ =
C_3 \wedge C_4 \wedge C_5$ where $C_3 = \overline{x_0} \vee
\overline{x_2} \vee x_4$, $C_4 = x_0 \vee \overline{x_4}$, $C_5 = x_2
\vee \overline{x_4}$, \bm{F_{g_2}}$ = C_6 \wedge C_7 \wedge C_8$ where
$C_6 = x_3 \vee x_4 \vee \overline{x_5}$, $C_7 = \overline{x_3} \vee
x_5$, $C_8 = \overline{x_4} \vee x_5$, \bm{F_{g_3}}$ = C_9 \wedge
C_{10}\wedge C_{11}$ where $C_9 = \overline{x_5} \vee \overline{x_6}
\vee x_7$, $C_{10} = x_5 \vee \overline{x_7}$, $C_{11} = x_6 \vee
\overline{x_7}$.

Consider the problem of taking a set of clauses $G$ out of \prob{X}{F}
where $x_i\!\in\!X,i\!=\!0,\dots,7$. Assume, for the sake of
simplicity, that $C_0,\dots,C_{11}$ are in $G$. Assume that this PQE
problem is solved by \Dpqe that derives regular D-sequents introduced
in~\cite{fmcad13}.  We also assume that \Dpqe is currently in subspace
$\pnt{r}= (x_6 = 0)$. Note that the gates $g_0,g_1,g_2$ are
``unobservable'' in this subspace. That is their values do not affect
the output of $M$ no matter what input \pnt{v} is applied (as long as
\pnt{v} produces the assignment $x_6=0$). Here \pnt{v} is a full
assignment to $V= \s{v_0,\dots,v_k}$. Let us show that after entering
the subspace \pnt{r}, \Dpqe derives atomic D-sequents
\oods{r}{C_i},$i=0,\dots,8$. These D-sequents express the
unobservability of $g_0,g_1,g_2$ in subspace \pnt{r}.

First consider the clauses of $F_{g_3}$.  Since $C_9$ is satisfied by
\pnt{r} and $C_{10}$ is implied by the clause $C_{11}$ in subspace
\pnt{r}, $C_9$ and $C_{10}$ are removed \prob{X}{F} as redundant in
subspace \pnt{r}. Now consider the clauses $C_6,C_7,C_8$ of
$F_{g_2}$. Since the clauses $C_9$ and $C_{10}$ are removed, the
clauses of $F_{g_2}$ are blocked at the variable $x_5$ in subspace
\pnt{r}. So, the D-sequents \oods{r}{C_i}, $i=6,7,8$ are derived.
Since the clauses of $F_{g_2}$ are removed from the formula in
subspace \pnt{r}, the clauses of $F_{g_0}$ and $F_{g_1}$ are blocked
at $x_3$ and $x_4$ respectively. So, the D-sequents \oods{r}{C_i},
$i=0,\dots,5$ are derived.

\end{document}